\documentclass[sigconf,nonacm]{acmart}
\usepackage{graphicx}

\usepackage[nolist]{acronym}
\acrodef{cs}[CS]{Computer Science}
\acrodef{cser}[CSER]{Computer Science Education Research}
\acrodef{it}[IT]{Information Technology}
\acrodef{sinz}[SINZ]{Software Innovation NZ}
\acrodef{qa}[QA]{Quality Assurance}
\acrodef{api}[APIs]{Application Programming Interfaces}
\acrodef{iac}[IaC] {Infrastructure-as-code}
\acrodef{wil}[WiL] {Work Integrated Learning}
\acrodef{slrs}[SLRs] {systematic literature reviews}

\AtBeginDocument{%
  \providecommand\BibTeX{{%
    \normalfont B\kern-0.5em{\scshape i\kern-0.25em b}\kern-0.8em\TeX}}}

\begin{document}

\title{From student to working professional: A graduate survey}

\author{Jacqueline Whalley}
\email{jacqueline.whalley@aut.ac.nz}
\orcid{0000-0001-7633-5200}
\affiliation{
\institution{Auckland University of Technology}
\city{Auckland}
\country{New Zealand}
}

\author{Asanthika Imbulpitiya}
\orcid{0000-0003-3730-7252}
\email{Asanthika.Imbulpitiya@op.ac.nz}
\affiliation{
\institution{Otago Polytechnic}
\city{Auckland}
\country{New Zealand}
}

\author{Tony Clear}
\email{tony.clear@aut.ac.nz}
\orcid{0000-0001-6985-5064}
\affiliation{
\institution{Auckland University of Technology}
\city{Auckland}
\country{New Zealand}
 }

\author{Harley Ogier}
\email{harley.ogier@lincoln.ac.nz}
\orcid{0000-0001-6985-5064}
\affiliation{
\institution{Lincoln University}
\city{Lincoln}
\state{Selwyn}
\country{New Zealand}
}

\renewcommand{\shortauthors}{Whalley et al.}

\begin{abstract}
This paper reports on the results of a 2023 survey that explores the \ac{wil} experiences of thirty recent \ac{cs} graduates. The graduates had all completed their undergraduate bachelors degree within the last five years and were currently employed in a \ac{cs} industry role. The survey asked about the graduates’ perceptions within a continuum of \ac{wil} experiences from final year capstone projects to professional development in their first industry-based role. Most respondents had taken a capstone course involving a team project. Only two respondents had participated in an internship program. Our results indicate that graduates value their capstone experiences and believe that they provide transferable skills including teamwork, managing client relations, exposure to technologies and methods, and time management. When entering their first industry role less than fifty percent of graduates were allocated a mentor. Overwhelmingly, these graduates noted the importance of those mentors in their transition from student to working professional. Very few of the surveyed graduates were provided with ongoing professional development opportunities. Those who did noted significant gains including growth of leadership skills and accelerated career progression. Our survey highlights a gap and an opportunity for tertiary institutions to work with industry to provide graduate onboarding and novice/early-career professional development opportunities. 
\end{abstract}

\ccsdesc[500]{Social and professional topics~Computing education}

\keywords{graduate survey, work integrated learning, capstone courses}

\maketitle
\pagestyle{empty}

\section{Introduction}
The Computing Curricula 2020 Task Force report highlighted the need for academic and industry partnerships to address the growing skills gap between CS/SE graduate skills and knowledge, and employer expectations \cite[p. 40]{CC2020}. The process of finding their first job for many graduates is fraught with difficulty. There is a divide between graduate skills and knowledge and what industry is expecting \cite{emsi,zdnet}. The tech industry for years has been reluctant to hire and invest in upskilling and developing graduates entering the industry \cite{matthews2021,CC2020, zdnet,employersClear2015}. So how does a new graduate bridge that gap? How do we, as academics and educational institutions, assist them to bridge that gap?

To date very little academic research, and in turn research informed work, has been undertaken to examine graduate onboarding models other than the traditional university capstone projects and internships. There is even less work evaluating the effectiveness of such initiatives. 

\section{Related Work}
In this section we present a high-level overview of the literature on capstone courses and graduate onboarding approaches. This reveals a continuum from university initiatives intended to prepare students for work, to industry-based opportunities for development as novice or early-career professionals.

\subsection{Academic Initiatives}
Most undergraduate bachelors degrees in \ac{cs} and \ac{it} have a compulsory course taken during the final year, that is considered to be a \textit{capstone} course. Such courses are intended to provide an integrative educational experience that ensures work readiness \cite{Clear2001CapstoneDefinition}. The classic capstone course can be viewed as a "half-way
house" with a combination of academic control and industry or workplace immersion. In addition to this classic view, several variants of capstone courses exist: \ac{wil} \cite{Clear2011WiL, Tappert2015ARC, Keogh2007}, internships \cite{Minnes2021}, sandwich programs \cite{Patel2012}, co-op education \cite{Clear2011WiL}, etc.  In contrast to  classic capstones, internships have little academic control and full workplace immersion. This research is focused on the experiences of graduates who completed a classic capstone  and/or an internship (aka work placement) during their degree.

The \ac{cser} literature presents a wealth of papers on capstone projects. However, over the last ten years most papers have focused on course design and curricula \cite{Davis2016TwoCapstones, martin2019designing, Allen2021DataScience}, use of Agile methods in team capstones \cite{Ding2017AgileScrum}, project selection \cite{Braught2022ProjectSelection}, and student factors \cite{Paasivaara2018Attitudes, Whalley2017Values, Bastarrica2017StudentViews}. Various \ac{slrs} detail the literature on capstone course design in the areas of Software Engineering \cite{TENHUNEN2023107191, Dugan2011}, \ac{it} \cite{martin2019designing} and assessment of capstone projects \cite{trevisan2006review}. One recent paper reported on inter-institutional trends in capstone projects from 2011 to 2018 \cite{capstoneTrendsBicol2011-2019}. Most published works in the area are framed as experience reports. Very few \ac{cser} articles on capstone courses are founded in empirical research and even fewer focus on attempting to assess the impact of capstone courses on graduate experiences as early-career professionals. 

\subsection{Onboarding Approaches}
Onboarding new employees has been an issue of longstanding concern for the tech industry, with \citet{onboardingSim1998} over twenty years ago noting the dominant approach was generally laissez-faire.  

In 2019 a repertory grid instrument \cite{repertoryGrid} was used to map the contributions of onboarding techniques to a set of onboarding goals derived from the literature \cite{agileOnboarding}.  As part of this study eleven developers who were new members joining  established agile software development teams were interviewed. Most of the developers interviewed were recent graduates. A broad range of industry-based onboarding techniques were reported, the most common being mentorship, peer support, and online communities. Mentorship experiences ranged from formal modes with regular meetings to informal modes with ad-hoc interactions or a mixture of both. In formal situations participants had the perception that as a mentee they could seek help as needed from their mentor even if they did not in practice. Peer support was informal and ad-hoc. For solutions to technical problems, the participants relied on activities like searching online communities (e.g. Stack Overflow, Quora, and MSDN) and reading internal technical documents. Only three of the new team members interviewed took part in a formal induction program. 

\citet{matthews2021} suggested, in 2021, that the most common current onboarding model is that of short-term graduate internships. However, he considered internships to be too selective and limited because they tended to be tailored to mainstream areas. Moreover, these industry based models are considered to typically be under resourced and restricted in scope \cite{matthews2021}. 

Learning on the job strategies, such as those adopted in the U.K. government sponsored university-level Graduate Apprenticeship scheme, have been said to have "dual underlying goals: degree apprenticeships are tasked with producing skilled employees and increasing social mobility \cite{SmithMobility2021}". A policy analysis by \cite{NawazImpact2023} covering  a range of U.K. industries has reported "a shortage of depth and breadth of concrete and peer-reviewed evidence on the impact of degree apprenticeships. Nevertheless, existing data demonstrate that degree apprenticeships are meeting their intended purpose of contributing positively to the UK Government’s high-level goals for productivity and social mobility".

However, U.K. degree apprenticeships have been criticised in \cite{LearningAgmtsBallew2018} for failing to even-handedly serve the needs of all stakeholders (students, academics, academic institutions, employers, parents, citizens and government) in the scheme. Smith et al. \cite{SmithMobility2021} have also critically observed, that from an IT industry perspective, "Apprentices were found to be drawn from all socioeconomic groups and represented those new to work and upskillers already in work. For upskillers, the degree apprenticeship offered a belated opportunity for degree-level study. However, young people recruited into the apprenticeship were disproportionally from more privileged groups". So the mobility goals of the scheme were not being achieved, even if some of the onboarding goals may have been met.

\section{Method}

\subsection{Participants}
\begin{figure}[h]
    \centering
    \includegraphics[width=1\columnwidth]{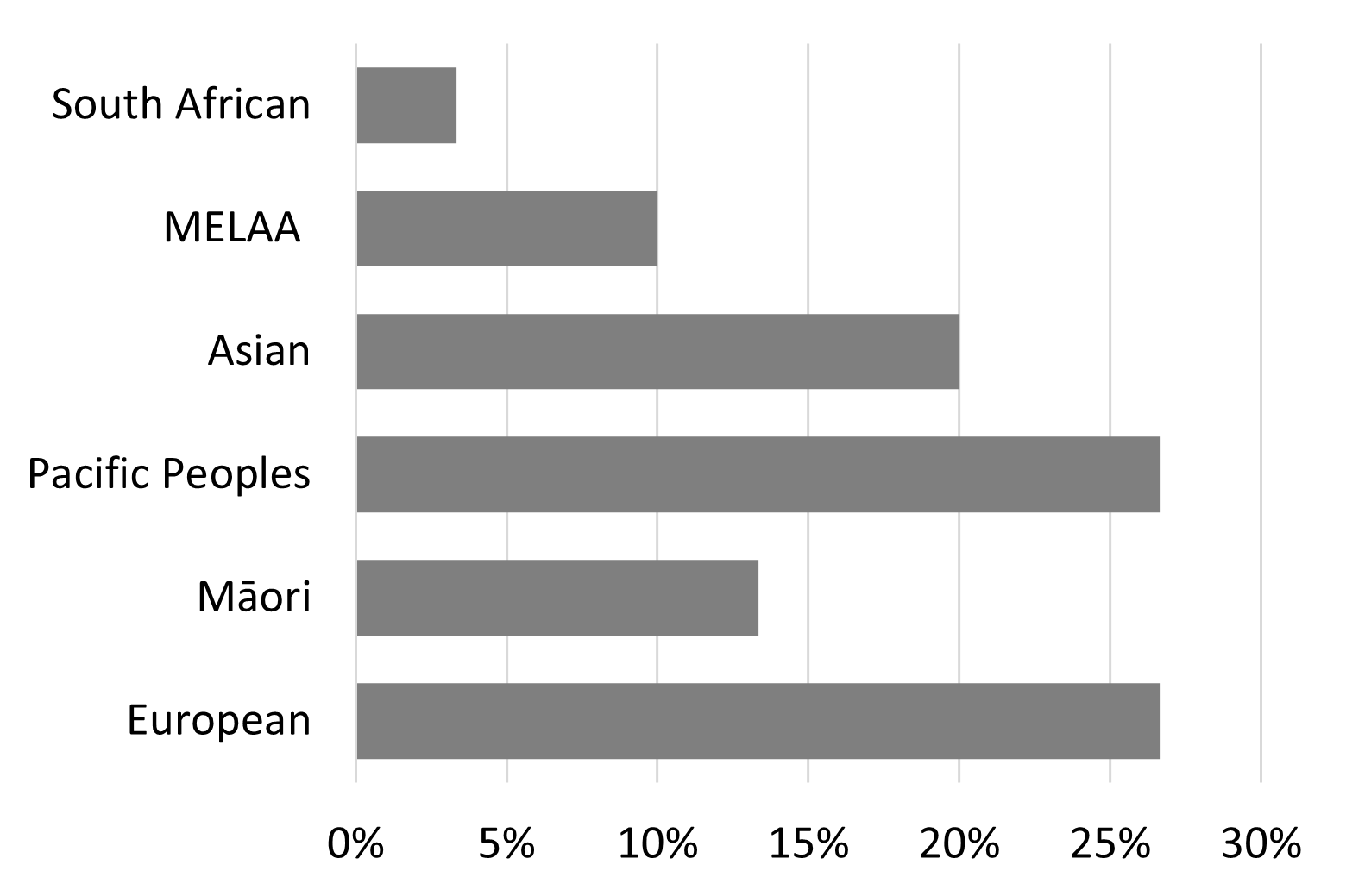}
    \caption{Participant ethnicity \protect\footnotemark}
    \label{fig:ethnicity}
\end{figure}

Our Participants were recent graduates (in the last five years) and  were all currently employed in a role in the tech sector. We recruited participants through local alumni LinkedIn pages and through \ac{sinz}: an organisation that links industry, academia and government agencies. Thirty respondents completed the survey. Fifty three percent of respondents reported identifying as male. The remaining 47\% reported identifying as female. No other gender identities were reported. The ethnic backgrounds of the participants were diverse (Figure~\ref{fig:ethnicity}), with the majority reporting as European (27\%) or Pasifika (27\%).

\footnotetext{MELAA encompasses those graduates identifying as Middle Eastern, Latin American or African}

\subsection{Survey}
The mixed design survey (see Appendix~\ref{appendix:survey}) included a mixture of open ended, Likert scale, nominal, and dichotomous questions. Nominal questions were comprised of a list of potential categories as response options. An "other" option was included to capture additional unforeseen data. Respondents could select one or more options. The survey was peer reviewed and refined in collaboration with four local capstone course instructors. The only demographic information collected was related to gender identity and ethnicity. Dichotomous (yes or no) questions were used to ensure subsequent questions were relevant to the experiences of the participant. The survey was was delivered online through the Qualtrics system for a period of two months. Geolocations were collected by default and Qualtrics was set to flag duplicate submissions.

\subsection{Analysis}
The Likert scale and binary data were analysed using simple descriptive statistics. A descriptive approach was taken to analysing the responses to open questions. Common views and experiences were identified and reported as domain summaries \cite{Boyatzis} and supported with participant quotes. Exceptions or idiographic views and experiences are also discussed because, while these were not recurrent patterns in the data, they represent valuable perspectives and provide further insight into the experiences of recent computer science graduates in their first professional role.

\section{Results}
\begin{figure}[h]
    \centering
    \includegraphics[width=1\columnwidth]{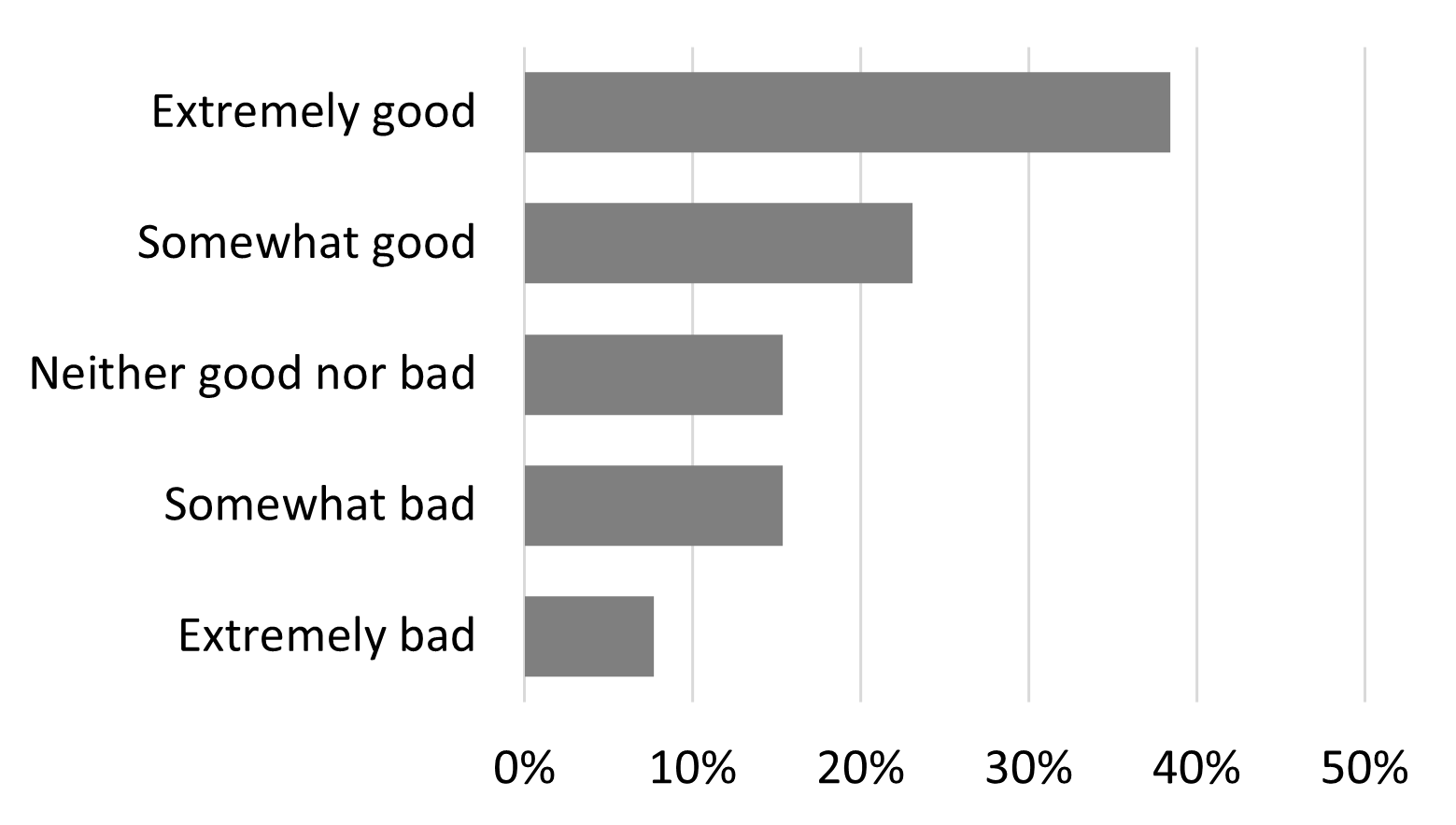}
    \caption{Capstone experience}
    \label{fig:capstoneExp}
\end{figure}

Twenty eight of the recent graduates had taken part in a capstone project as part of their undergraduate bachelors degree program.  Figure~\ref{fig:capstoneExp} shows their responses to Q1, which asked what their experience had been during their project. The remaining two respondents had been involved in an internship program as part of their degree.

\subsection{Capstone Projects}
The 28 respondents who had completed a capstone project as part of their degree were asked if it helped their growth and their transition into an industry role. Only two of the respondents said that their capstone projects did not help their growth and their transition into industry. Of those, one mentioned that they were employed in a networking role and that this was not their major. The other mentioned that the stress of undertaking an industry based capstone project online during the pandemic, as well as the lack of commitment of team members, resulted in an experience that they perceived as being of little value. One graduate stated that their capstone was of use but that they had learnt processes which they had not used in practice. As they already worked in industry and had been studying part-time they felt their capstone experience did not result in meaningful professional or discipline knowledge growth. Other respondents included aspects of their learning and growth which helped or applied to their current positions in terms of:\\

\noindent\textbf{Teamwork}
\begin{quote}\textit{``Yes, [it] helped me understand how IT projects work, how to coordinate and work within a project team."}\end{quote} 

\begin{quote}\textit{``Yes, there were group dynamics which more closely resembled a real-world team than other uni group projects. The work itself was almost identical to a real-world project, but on a longer timeline to allow for learning.''}\end{quote}
    
\noindent \textbf{Client Relations}

   \begin{quote} \textit{``There were client disruptions and issues which was a great reflection of what graduates will see in the industry''}\end{quote}
    
\noindent \textbf{Technologies}

    \begin{quote}\textit{``The project we took on was probably my first exposure to [GitHub] and working on a single code base as a group of individuals. While I am more involved in platform engineering now, learning how to contribute to live code was extremely useful for the early stages of my career.''}\end{quote}

\noindent \textbf{Methods} 

    \begin{quote}\textit{``Yeah, I got to experience first hand the agile methodologies. We somewhat didn't stick to its principles. However, it taught me things around communicating and being comfortable about it. It further highlighted the importance of user testing to make sure we are not building something that our client won't be using. Even though it would've been "cool" for us devs but it's wasted effort if our consumers won't use it.''}\end{quote}

\noindent \textbf{Time management}

    \begin{quote}\textit{``Yes, because I've witnessed this one thing that time is everything when working in industries with deadlines of tasks to complete.''} \end{quote}
    
    \begin{quote}\textit{``Absolutely, it has given me the mindset I need to be able to accomplish goals within a reasonable time frame.''}\end{quote}
    
Only one graduate discussed their capstone experience as being central to building leadership skills and initiative.
\subsection{Internships}
Two of the graduates undertook internships rather than a capstone project course. One was required to find and organise their own internship while the other was supported by their institution to find their work placement. Both reported having a good experience on their internship. One described joining a  specialist equipment company that produced software as a \ac{qa} tester. They created \ac{api} for test reports. They mentioned that \textit{``Overall I wasn't a huge fan of QA testing (quite repetitive) but helped me realise I did enjoy doing development, and could do it 40hrs/week.''}. The other mentioned that their internship was of short duration but that it was "\textit{a great learning opportunity and added variety to me [sic] degree experience''}.

\subsection{Industry Graduate Support}
Forty-seven percent (14) of our respondents were allocated a mentor when entering their first graduate position in industry. All had positive experiences with over 80\% rating their employers' mentorship program as extremely helpful. Those graduates who did have mentors highlighted a number of benefits with respect to their mentors enabling their integration into a professional environment.

\begin{quote}\textit{``Guidance, understanding, empathy, a touch point for advice and also a voice for me to help me get noticed and start to have my own voice''}\end{quote}

\begin{quote}\textit{``My mentor was very helpful in guiding me down the right path. I got to learn off of his experiences. Can't beat the experience you get to feed [off] your mentor''}\end{quote}

Some respondents highlighted their mentor's role in their ongoing learning and workload management:
\begin{quote}\textit{``My mentors filled in all the gaps of my knowledge and gave me digestible tasks to carry out while I integrated.''}\end{quote}

\begin{quote}\textit{``I joined a team that was just starting its first IaC [Infrastructure as Code] project and was mentored under one of the contractors brought in for the project. He was super helpful in getting me set up on Terraform, and helped me understand IaC as a whole within the Azure cloud space''}\end{quote}

For those who were not allocated a mentor we asked them about how hard they perceived their transition into their first industry role.

One graduate commented on having to direct their own learning in order to do the job. They \textit{``Researched, video tutorials and books, and LOTS of trial and error based of feedback, while carefully balancing the impression of expertise from the clients perspective.''} This experience of learning on the job without support was shared by other respondents: \textit{``I just learnt on the job, wasn't exactly able to ask anyone for help in my specific line of work''}. 

While some recent graduates perceived not having a direct mentor as a disadvantage, others did not.
\begin{quote} \textit{``I am quite self-sufficient and coped well, and there were senior members of staff available to me for questions and training. I think i benefited more from having a diverse cast of people to learn from than a single mentor.'' }\end{quote}

Only 15\% of the graduates surveyed undertook a formal induction course, lasting from two months to eighteen months, on entering their first job. Of the graduates who underwent an induction program, all were offered that program in-house. It should be noted that in New Zealand there is a prevalence of small businesses with less than 20 employees (97\% of all firms contributing over a quarter of  New Zealand’s gross domestic product). Many tech companies are considered open source companies. Consequently, many \ac{cs} graduates start their careers in relatively small companies working with open source software and \ac{api}. This may be one reason why in our survey such a small number reported participating in a formal induction program and why so few had the opportunity to build knowledge through formal training offered in-house or through software vendors. 

\subsection{Ongoing Development}
Finally, we asked the graduates surveyed about the ongoing professional development opportunities that they had been offered by their employer since leaving university. One respondent noted that they had no opportunities. One graduate said that they have had to take ownership of their own development and have chosen to take online courses to further their learning in their own time. One respondent mentioned having had some very limited opportunities to undertake Pluralsight \cite{Puralsight} courses. While a wide variety of opportunities were noted, respondents typically only had access to one form of professional development. Among the opportunities noted was a common theme of being given the opportunity to explore other roles in the company:

\begin{quote}\textit{``I didn't start in a \ac{it} role but they were aware I studied it and had \ac{it} experience. They have given me many opportunities to grow in the \ac{it} space with my first role as \ac{it} project coordinator and then moving to \ac{it} project manager.''}\end{quote}

\begin{quote}\textit{``Being given [the chance] to explore  other responsibilities outside of my main purpose. Things like BA or PO work. It really helped with my negotiation, communication, and explain[ing] technical context in simpler terms that every[one] can understand.''}\end{quote}

Others noted specific opportunities to develop their knowledge:

\begin{quote}\textit{``I’ve been doing cloud security courses and learning more about AD [Active Directory] Risks, been taking trips around the country to set networks up at `Company A',\protect\footnotemark branches.''}\end{quote}

\begin{quote}\textit{``Access to conferences, with recorded material. Weekly half day to pursue passion projects and sharing team learnings. Flat hierarchy with ample role growth opportunities.''}\end{quote}

\footnotetext{Note businesses that the participants currently or have previously worked for have been anonymized herein}

\begin{quote}\textit{```Company B', and my current company, have been excellent in the space for growth. I work in the Azure space, and they have always encouraged going to the instructor led courses Microsoft offers. These are normally 3-4 full day sessions ...''}\end{quote}

Often software vendors will provide the training needed to work within their ecosystems. It is interesting that only one of the graduates mentioned training of this nature.

One respondent mentioned that their company provided financial support for professional certifications associated with their current role.

Most respondents who had access to further professional development reported that these industry based development opportunities were very or extremely useful. However, it is important to note that over two thirds of graduates in our survey did not have access to professional development opportunities. The reported benefits of such opportunities included: 
\begin{itemize}
    \item improved technical skills
    \item growth of leadership and management skills
    \item ability to reach goals/targets and experience more ``wins''
    \item accelerated career progression
    \item broadening of employment opportunities with ability to move into different roles or areas within the tech sector
    
\end{itemize}

\section{Discussion}
The facets of the capstone which graduates perceived as most applicable to the competencies required in their professional roles were associated with "professional knowledge areas and dispositions" \cite{CC2020} such as the ability to work and coordinate activities within a team, manage client relationships, and manage one's time. The technical skills that were viewed as transferable were more related to processes and practices than specific technologies. Similar findings were noted by \citet{Bastarrica2017StudentViews} who reported that by the end of their capstone student perceptions of the value of professional skills had increased while the perceived value of technical skills had decreased. 

The results of our survey suggest that capstone courses are indeed providing graduates with some of the competencies that facilitate their transition into work. However, there is still a mismatch between the skills acquired at university and those required in a graduate's first industry role. The technical skills taught at university are almost inevitably not the ones graduates will need in their early career due to the rapid pace of change of technologies. The need to have internalised a set of dispositions such as \emph{being adaptable, self-directed and proactive} {\cite{CC2020, Frezza2020}} is key.  Being able to demonstrate these dispositions in a work setting is vital. In a recent report on the digital skills shortage in New Zealand \cite{DigitalSkillsNZ2021}, one \ac{it} company CEO stated that they struggled to find candidates with critical problem solving, communication, and negotiation competencies. However, the report also noted that companies ``tended to recruit externally for new skills, rather than develop their own workforce.'' \cite[p. 45]{DigitalSkillsNZ2021}. Moreover, employers found it a challenge to find time for the development of employee competencies, instead prioritising business-as-usual activities. This is reflected in our findings where most of the graduates surveyed invested their personal time into professional development with little or no employer support. 

Supporting graduates' transition to professional work is central to making them employable and productive \cite{Chong2014Transition}. Alongside this, tertiary institutions have been and continue to  experience pressure from government and industry to produce employable graduates \cite{Bridgstock2019}. 

Our results point to some areas of concern for employers too. In a study of onboarding success, organisational fit, and turnover intentions, \citet{Sharma2020} found that \emph{"Providing ongoing support to new recruits so that they feel supported in
their job is likely to be most important in getting new staff to settle in"}. Consequently, new recruits who do not feel supported were deemed more likely to leave the organisation. Given that less than half of our graduates were allocated a mentor, it appears that provision of ongoing support was often lacking, placing both employers and employees at risk of dissatisfaction. This points to the critical role of \emph{mentorship} as also noted in the Master's study by He (cited in \cite{Clear2023Dispositions}).

\section{Conclusion}
The results of a survey of the Work Integrated Learning (WiL) experiences of thirty recent New Zealand Computer Science (CS) graduates have been reported here. Key insights show that most graduates participated in a classic capstone course, which appears to serve as the primary means employed by institutions to provide students a bridge to their professional careers. Our results indicate that graduates value their capstone experiences and believe that they provide them with a set of transferable competencies including teamwork, managing client relations, exposure to technologies and methods, and time management.

When entering their first industry role less than fifty percent of graduates were allocated a mentor. Overwhelmingly, these graduates noted the importance of those mentors in their transition from student to working professional. Very few of the surveyed graduates were provided with ongoing professional development opportunities. Those who did noted significant gains including growth of leadership skills and accelerated career progression. Our survey highlights a gap in support around the transition from academia to industry, with risks to employing organisations and graduates lacking support in their early careers. This provides an opportunity for tertiary institutions to work with industry to provide graduate onboarding, in addition to novice and early-career professional development opportunities.

Our study was conducted in the context of a relatively small country, where the tech sector is dominated by small to medium-sized enterprises (SMEs). The experiences of graduates entering the tech sector in other contexts, particularly those with a greater proportion of large-scale enterprises, may differ from those reported here and are worth investigating.



\begin{acks}
\emph{[Redacted for review.]}
\end{acks}

\appendix
\section{Survey}
\label{appendix:survey}
\textbf{--Capstone Project}

 \noindent\textbf{Q1}. Did you complete a final year capstone project as a part of your degree? Yes | No

\noindent\textbf{Q1.1}. How was your experience in the capstone project as an undergraduate? 5pt Likert (Extremely Bad--Extremely Good)

\noindent\textbf{Q1.2}. Did your capstone project help your growth and transition to industry? If so, how?

\noindent\textbf{--Internship}

\noindent\textbf{Q3}. Did you complete an internship or a work placement as an undergraduate? Yes | No

\noindent\textbf{Q3.1}. How did you find your work placement or internship? University assisted me in finding an opportunity | I was required to identify an opportunity myself | Other

\noindent\textbf{Q3.2}. In what way did the university assist you in finding an internship or a work placement?

\noindent\textbf{Q3.3}. How was your experience in the internship/work placement as an undergraduate? 5pt Likert (Extremely Bad--Extremely Good)

\noindent\textbf{Q3.4}. Did this help your growth in transitioning to the industry? If so, how?

\noindent\textbf{--Workplace Mentorship}

\noindent\textbf{Q4}. As a new graduate, were you allocated a mentor when you first started working? Yes | No

\noindent\textbf{Q4.1a}. How useful did you find this initiative? 5pt Likert (Not at all useful--Extremely useful)

\noindent\textbf{Q4.2a}. What aspects were useful, and how did your mentor help you to integrate into this professional environment?

\noindent\textbf{Q4.1b}.How did you cope without having a mentor when you first started working?

\noindent\textbf{--Workplace Professional Development}

\noindent\textbf{Q5}. As a new graduate did you participate in an induction program? Yes | No

\noindent\textbf{Q5.1}. Was this an in-house internal program or was it offered by a third party? In-house | Third party

\noindent\textbf{Q5.2}.How long did the program last?

\noindent\textbf{Q6}. What other opportunities has your employer provided in terms of professional development since you joined?

\noindent\textbf{Q7}. How useful were these opportunities? 5pt Likert (Not at all useful--Extremely useful)

\noindent\textbf{Q8}. In what way did these opportunities assist in your development as a professional?

\balance

\bibliographystyle{ACM-Reference-Format}

\bibliography{bib.bib}

\end{document}